\documentclass[printnumbers,amsmath,amssymb,twocolumn]{revtex4}
\usepackage{graphicx}
\usepackage{color}
\usepackage{float}

\begin{document}

\title{Density of States below the First Sound Mode in 3D  Glasses}

\author{Lijin Wang$^{1}$, Licun Fu$^{1}$, and Yunhuan Nie$^{2}$ \\
$^{1}${School of Physics and Optoelectronic Engineering,  Anhui University, Hefei 230601, China}\\
$^{2}${Beijing Computational Science Research Center, Beijing 100193, China}\\
}

\begin{abstract}

Glasses feature universally low-frequency excess vibrational modes beyond Debye prediction, which could help rationalize, e.g., the glasses' unusual temperature dependence of thermal properties compared to crystalline solids. The way the density of states of these low-frequency excess modes $D(\omega)$ depends on the frequency $\omega$  has been debated for decades.  Recent simulation studies of 3D glasses suggest that $D(\omega)$ scales universally  with $\omega^4$ in a low-frequency regime below the first sound mode. However,  no simulation study has ever probed as low frequencies as possible to test directly whether this quartic law could work all the way to extremely low frequencies.  Here, we calculated $D(\omega)$  below the first sound mode in  3D glasses over a wide range of frequencies. We find  $D(\omega)$ scales with $\omega^{\beta}$ with ${\beta}<4$ at very low frequencies examined, while the  $\omega^4$ law  works only in a limited intermediate-frequency regime in some glasses. Moreover, our further analysis suggests our observation  does not depend on glass models or  glass stabilities examined. The $\omega^4$ law of $D(\omega)$ below the first sound mode is dominant in current simulation studies of 3D glasses,  and our direct  observation of the breakdown of the quartic law at  very low frequencies thus leaves an open but important question that  may attract more future numerical  and theoretical studies.

\end{abstract}

\date{\today}

\maketitle

\section{Introduction}

The disorder  endows  glasses with special vibrational properties that are different from those of their crystalline counterparts.  In crystals, vibrational  modes at low frequencies $\omega$ are plane waves (phonons) and their density of states  could be described perfectly by Debye theory~\cite{kittel}, while  there are low-frequency vibrational modes in excess of the Debye prediction in glasses~\cite{parisi_nature,tanaka_nm2008,Buchenau2007,mizuno_pnas,Wang2019NC,Wang2021prl,spinglassPRL}. These excess modes, termed usually  non-phononic or quasi-localized modes~\cite{Xu2010EPL}, could be observed in a broad class of glasses, and hence has become one universal hallmark of glasses. Importantly, it has been suggested  explicitly that  these  excess  modes could also give some insight into the understanding of glasses' other elusive properties~\cite{manning_prl,Widmer-Cooper_np,chen_prl,rottler_prx,Zylberg_PNAS2017,Wang_prl2014,Flenner_SM2020,
Wang2019SMattenuation,Xu2010EPL,Wang_softmatter2012}, e.g., the mechanical and thermal properties of glasses, and the dynamics of supercooled liquids.

One essential question with respect to the low-frequency excess modes in glasses  is how the density of these  modes $D(\omega)$ scales
with $\omega$,  which has been a puzzle for decades and is now still a subject of  active research.
There have been some phenomenological models and theories~\cite{Shimada2018pre,Bouchbinder2020,Stanifer2018pre,Ji2019pre,Ikeda2019pre,meanfield1,meanfield2,Buchenau1991,Schober1996,Gurevich_prb2003,Gurarie_prb2003,Kumar2021arxiv,Schrimacher_prl2007,Xu2017prl,scipost2022,patrick} which have made  predictions with respect to  the  value of the exponent ${\beta}$ in $D(\omega) \sim \omega^{\beta}$. A compilation of related studies suggests ${\beta}$ could take a value of $2$, $3$ or $4$ in glasses under different circumstances; yet sometimes different values of ${\beta}$  even for the same glasses were predicted basing on different theoretical frameworks.

Nowadays, computer simulations are playing important roles in revealing how $D(\omega)$ evolves with $\omega$, which has attracted considerable attention~\cite{spinglassPRL,mizuno_pnas,Wang2019NC,Wang2021prl,ikeda_pre2018,Shimada2020pre,lerner_prl2016,lerner2018jcp,Lerner2020pre,Angelani2018,Lerner2020pnas,Bonfanti2020prl,Richard2020prl,Das2021prl,
 LernerPRE2017,Ji2020pre,Lerner2Dprl,Krishnan2D,Parisi2019prl,Gartner2016Scipost,Kapteijns2020pre,Giannini2021pre}  in the past less than one decade. We divide recent numerical  studies of the excess modes in structural glasses into two groups according to the system sizes examined. The first  group~\cite{mizuno_pnas,Wang2019NC,Wang2021prl,ikeda_pre2018,Shimada2020pre} studied  very large systems. However, it's notoriously known that it's not an easy task to calculate  the density of the low-frequency excess modes in large systems,  because extended (phonon-like) modes and excess modes could hybridize at low frequencies in these systems. Recently, two kinds of methods which we refer to  as the direct method and  indirect method,  have been proposed to calculate the density of excess modes in large systems. One  direct method refers to classifying  extended and excess modes using an order parameter~\cite{mizuno_pnas}, e.g., participation ratio, and then calculating the density of these excess modes; in other direct methods, it was proposed to use the  nonlinear  modes~\cite{Gartner2016Scipost,Kapteijns2020pre} that do not hybridize with extended modes even in very large systems, to represent excess modes, whereas it was reported subsequently a bond-space operator ~\cite{Giannini2021pre}  could do better  in disentangling excess and extended modes. For the indirect method, it refers to, e.g., subtracting the Debye contribution from the total density of states~\cite{Wang2021prl,Schrimacher_prl2007}. Recent studies show that the direct  and indirect methods  could give consistent results in 3D large glasses,  but lead to conflicting conclusions in  2D ones. Specifically, in large 3D  glasses, it was observed $D(\omega) \sim \omega^{4}$ using  direct and  indirect methods~\cite{mizuno_pnas,Wang2019NC}; however, in large 2D glasses, Mizuno, Shibab, and Ikeda~\cite{mizuno_pnas} found there is very few to even no excess modes using one direct method, while Wang, Flenner, and Szamel~\cite{Wang2019NC} observed a low-frequency regime where $D(\omega) \sim \omega^{2}$ using one indirect method.

The second group~\cite{lerner_prl2016,lerner2018jcp,Lerner2020pre,Angelani2018,Lerner2020pnas,Bonfanti2020prl,Das2021prl,
 LernerPRE2017,Ji2020pre,Lerner2Dprl,Krishnan2D,Richard2020prl,Parisi2019prl} studied the low-frequency modes below the first sound modes in sufficiently small systems.
When the system size $N$ is smaller, the first sound mode frequency $\omega_1 = \sqrt{G/\rho} \left(2 \pi/L\right)$ ($G$ is the shear modulus, $\rho$ is the density, and $L$ is the box length) is pushed to a higher frequency, hence leaving  below $\omega_1$ a wider low-frequency window  where vibrational modes are  excess modes beyond Debye prediction.  Since excess modes  do not hybridize  with extended modes below $\omega_1$ in the context of very small systems, it's reasonable to assume the density of these excess modes is equal to the total density of states.  For this reason, Lerner, D\"{u}ring and Bouchbinder~\cite{lerner_prl2016} proposed  to study the density of   low-frequency modes below $\omega_1$  in small structural glasses, which ignites considerable studies.

 A compilation of  recent studies~\cite{lerner_prl2016,lerner2018jcp,Lerner2020pre,Lerner2020pnas,Bonfanti2020prl,Das2021prl,Lerner2Dprl,Richard2020prl} following the procedure  of studying small systems suggests that $D(\omega)\sim \omega^{\beta}$ with ${\beta}=4$  at low frequencies below $\omega_1$ in 3D glasses. We notice  a constraint was  made in a  recent numerical study~\cite{Lerner2020pre} to guarantee the universality of ${\beta}=4$ in 3D glasses. Specifically, it was argued that  ${\beta}=4$ is valid  only in  glasses which are sufficiently  small but larger than a characteristic system size, because ${\beta}$ suffers from finite-size effects~\cite{Lerner2020pre}.
Nevertheless, to the best of our knowledge, no simulation study  has ever tested directly whether the validity of the $\omega^4$ law could extend to very low frequencies  far below $\omega_1$ in 3D glasses, though $D(\omega)$  has recently been  assumed implicitly or explicitly to grow as $\omega^4$  from zero frequency~\cite{Lerner2020pre}.

 In this work, we examined the  low-frequency density of excess modes  below the first sound mode in 3D small glasses.
  We observed  the value of ${\beta}$ in $D(\omega)\sim \omega^{\beta}$ is  smaller than $4$ in  the lowest-frequency regime examined, and the previously reported  ${\beta}= 4$  works only in  an intermediate-frequency regime in some glasses. Our observations are  supported by  the analysis in three vastly different model glass formers, and in two glasses with different glass stabilities.

\section{Models and Methods}\label{simulation}

 We performed three-dimensional computer simulations  in  three  different glass models:

 \noindent (I) A binary  mixture with the harmonic potential, which is referred to as HARM model. A  detailed description of this model could be found in Ref.~\cite{Wang_softmatter2012}.  The volume fraction we used in this model is 1.5, and  the corresponding number density $\rho=N/L^3 \approx 1.53$ with $N$ system size.

 \noindent (II) A typical binary system with the  Kob-Anderson Lennard-Jones potential~\cite{KALJ},  KALJ model. We used $\rho=1.2$, and our  simulation details in this model glass could be found in Ref.~\cite{Wang_prl2014}.

 \noindent (III) A binary system with the inverse power law potential,  IPL model. We  followed the simulation details as described in Refs.~\cite{lerner2018jcp,Lerner2020pre} when studying  this model glass.

Our zero-temperature ($T=0$) glasses were created by performing a very rapid quench of  equilibrated liquid states at very high temperatures to $T=0$ via energy minimization.  Our energy minimization method used is the fast inertial relaxation engine method~\cite{fire} in HARM model,  but the conjugate gradient algorithm in both  KALJ and IPL models; we used the two different popular minimization methods to exclude the possibility that the main conclusion in this work is resulting from one specific minimization method.    Glasses generated by a very rapid quench from liquid states are very poorly annealed with small glass stability. In addition, we also generated annealed KALJ glasses with larger stability. Specifically, we  annealed  KALJ glasses for a time of $t_A=1000$ (in KALJ time units~\cite{KALJ}) at $T=0.1$ below the  glass transition temperature $T_g\approx0.3$~\cite{Wang_prl2014} before performing  an instantaneous quench to $T=0$. We note here the stability of the annealed  glasses could be determined by both $t_A$ and $T$ according to Ref.~\cite{Wang2020softmater}, and we studied only one combination of  ($t_A$, $T$) in this work.

We obtained the low-frequency normal modes of $T=0$ glasses by the diagonalization  of  the dynamic matrix using the ARPACK program package~\cite{arpack}. The density of states  reads
$D(\omega)=\frac{1}{3N-3} \sum_{l=1}^{3N-3} \delta(\omega-\omega_l)$ with $\omega_{l}$ the frequency of mode $l$. Debye theory predicts phonon modes in a finite system occur  at discrete frequencies.  Hence, an inappropriate choice of the bin size  may result in an inaccurate calculation of $D(\omega)$ around a phonon mode~\cite{Wang2021prl,Wang2019NC}. To avoid this issue, following the latest studies~\cite{Lerner2020pnas,Wang2021prl,Shimada2020pre,Parisi2019prl}, we calculated  the cumulative density of states  $I(\omega)=\int_{0}^{\omega}D(\omega')d\omega'$.  The bin size in $I(\omega)$ is not a divisor as in $D(\omega)$, which makes the numerical calculation of $I(\omega)$ free from the errors due to the choice of bin sizes. A very stringent and convincing  way  to check a  scaling of $I(\omega) \sim \omega^{\gamma}$  is to plot $I(\omega)/\omega^\gamma$ against $\omega$, and hence  we show most of our results using the $I(\omega)/\omega^5$ plot to test the $I(\omega) \sim \omega^5$ scaling. To characterize the extent of the mode localization,  we  calculated the participation ratio $p(\omega_l)=\frac{\left(\sum_{i=1}^N |\vec{e}_{l,i}|^2\right)^2}{N\sum_{i=1}^N |\vec{e}_{l,i}|^4}$   with $\vec{e}_{l,i}$ the polarization vector of particle $i$ in  mode $l$. A smaller value of participation ratio usually indicates a more localized mode.

\begin{figure}
\centering
\includegraphics[width=0.38\textwidth]{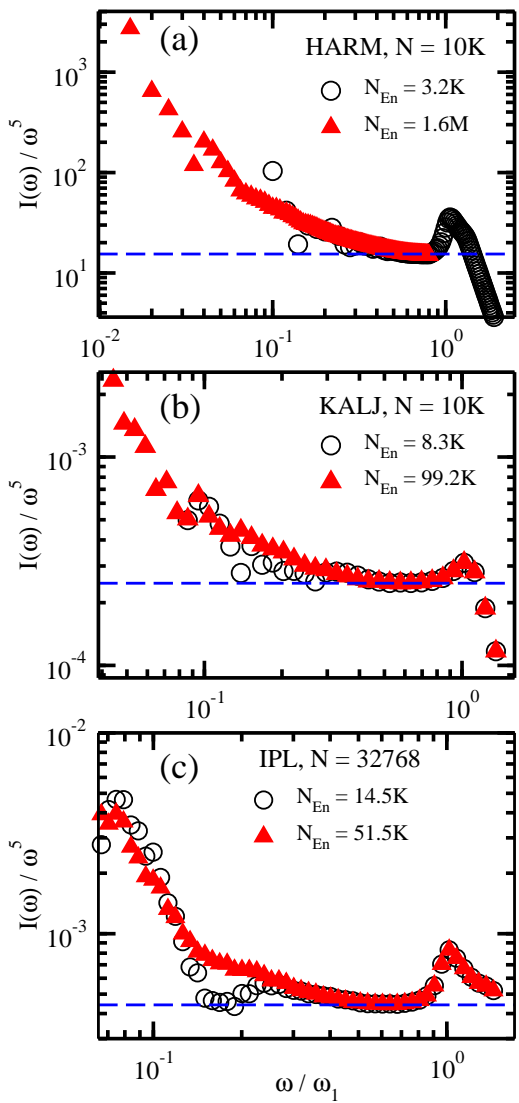}

\caption{\label{fig1} The ensemble size $N_{En}$ dependence of the reduced cumulative density of states  $I(\omega)/\omega^5$ against $\omega/\omega_1$  in three  different 3D model  glasses: system size $N=10K$ in  HARM model in (a), $N=10K$ in KALJ model in (b), and $N=32768$ in IPL model in (c). The first sound mode frequency  $\omega_1\approx 0.10$ in (a), $\omega_1\approx 1.11$ in (b), and $\omega_1\approx 0.72$ in (c). Glasses in (a), (b) and (c)  were  all created by a very rapid quench of very high temperature liquids to zero temperature.
 In all model glasses examined, one can see the breakdown of the
$I(\omega) \sim \omega^5$ law in the very low-frequency region.   The horizontal dashed blue line is a guide to eyes.
 }
\end{figure}

\begin{figure}
\centering
\includegraphics[width=0.4\textwidth]{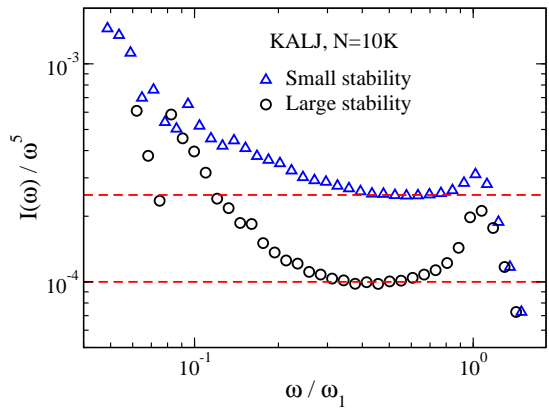}
\caption{\label{fig2} The glass stability dependence of the reduced cumulative density of states  $I(\omega)/\omega^5$ against $\omega/\omega_1$ in the KALJ model systems in 3D. The glasses with small stability refer to those generated by a very rapid quench to $T=0$ from  very high-temperature liquids, while the ones with large stability refer to annealed glasses as described in the main text. The first sound mode frequency $\omega_1\approx 1.11$ and $\omega_1\approx 1.16$ in glasses with small stability and large stability, respectively. The horizontal dashed  lines are a guide to eyes.
}
\end{figure}

\section{Results}

 Figure~\ref{fig1} shows  the reduced cumulative density of states  $I(\omega)/\omega^5$  in different 3D model glasses: HARM model in (a),
 KALJ model in (b), and IPL model in (c). Our  system size in each model  is chosen such that a visible bump in  $I(\omega) /\omega^5$  could be observed around the first sound mode frequency  $\omega_1$, and we focused mainly on the low-frequency modes below $\omega_1$. In the HARM model glasses, Fig.~\ref{fig1}(a),  one can see a short-range  plateau in the $I(\omega)/\omega^5$ plot in an intermediate-frequency regime, suggesting $I(\omega)\sim \omega^5$, i.e., $D(\omega)\sim \omega^4$, which is consistent with previous numerical results~\cite{lerner_prl2016,lerner2018jcp,Lerner2020pre,Lerner2020pnas,Bonfanti2020prl,Das2021prl,Lerner2Dprl,Richard2020prl}. However,  an obvious upturn in the $I(\omega)/\omega^5$  plot shows up when going to much  lower frequencies,
 suggesting the $I(\omega)\sim \omega^5$ scaling breaks down at very low frequencies, which is our main conclusion in this work.
 One can draw the same conclusion as in HARM model glasses after checking the results in KALJ model glasses in Fig.~\ref{fig1}(b)  and IPL model glasses in Fig.~\ref{fig1}(c).
 In computer simulations, one can probe lower frequencies by using a larger ensemble size (number of configurations) $N_{En}$.  We find that the observed low-frequency breakdown of the $I(\omega)\sim \omega^5$ law does not depend on $N_{En}$ examined, e.g., the observation remains almost unchanged with
  $N_{En}$ ranging from thousands to millions in Fig.~\ref{fig1}(a).

  We note here all glasses examined in Fig.~\ref{fig1} were created  by quenching instantaneously  high-temperature liquid states to $T=0$,  and we refer these poorly annealed glasses to as glasses with small stability. We then go a little further to check whether the observed breakdown of the $I(\omega)\sim \omega^5$ scaling   will persist in glasses with  larger  stability.
  In Fig.~\ref{fig2}, we compared $I(\omega)\sim \omega^5$ in two KALJ model glasses with different glass stabilities. Glasses with large stability here refer to  annealed KALJ model glasses as described in \emph{Section}~\ref{simulation}. Interestingly, the very low-frequency breakdown  of the  $I(\omega)\sim \omega^5$ scaling  could
  also be observed in glasses with larger stability.
   Moreover, one can observe the height of the intermediate-frequency plateau  in the $I(\omega)/\omega^5$ plot in glasses with large stability goes down  as compared with that in  glasses with small stability.
  This is consistent with previous studies~\cite{Wang2019NC,Lerner2020pnas,lerner2018jcp,Ji2020pre}  that suggest there are less excess modes in  more stable glasses. However, we note  the stability of our annealed KALJ glasses, though larger than that of our poorly annealed ones,  is still small with respect to ultra-stable glasses generated using the swap Monte Carlo method~\cite{Grigera2001,berthier_prx2017,berthier_prl2016}. Therefore, we cannot exclude the possibility that a very large enhancement of the glass stability may reach a conclusion  different from ours.

\begin{figure}[t]
\includegraphics[width=0.4\textwidth]{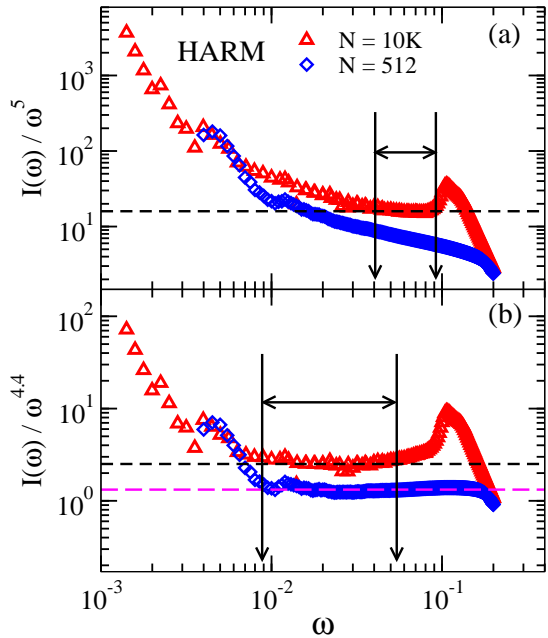}
\caption{\label{fig3}
System size $N$ dependence of the reduced cumulative density of states  $I(\omega)/\omega^5$ in (a) and  $I(\omega)/\omega^{4.4}$ in (b) in HARM
 model glasses in 3D. For reference, the first sound mode frequency $\omega_1\approx 0.10$ in $N=10K$, while $\omega_1\approx 0.27$ in $N=512$ glasses, which is beyond our largest frequency examined. One can observe in (a) there is an overlapping
frequency range (indicated by the horizontal line with arrows in both ends)
 where $I(\omega)/\omega^5$  in $N=10K$ is a plateau while $I(\omega)/\omega^5$ in $N=512$
increases with decreasing frequencies. In (b) we
indicated, in the same way as in (a),  another  overlapping frequency range  where $I(\omega)/\omega^{4.4}$ is a plateau in both  $N=10K$ and $N=512$ systems.
   The ensemble size  $N_{En}\approx3.19M$ in the $N=512$ system, and the $I(\omega)$ data in $N=10K$
    is composed of    $I(\omega)$  with $N_{En}=3.2K$ and  $N_{En}=1.6M$.
    For visualization purpose,  the vertical axis $I(\omega)$ in $N=512$ has been scaled  with a factor of  $0.382$. The horizontal dashed  lines are a guide to eyes.
 }
\end{figure}

Recent simulation studies~\cite{LernerPRE2017,Angelani2018,Lerner2020pre,Parisi2019prl} of the low-frequency modes below the first sound mode show that $\beta$  in $D(\omega) \sim \omega^{\beta}$ could be much smaller than 4, especially in  poorly  annealed 3D glasses.
Very recently, Lerner~\cite{Lerner2020pre}  attributes the $\beta < 4$  to  finite-size effects. Specifically, there is a crossover  system size $N_c$ determined from a core size   of soft quasi-localized modes; it was claimed ${\beta}=4$  as long as  the glass system size $N$ is  larger than  $N_c$, while $\beta < 4$ when $N < N_c$.  It was also shown in Ref.~\cite{Lerner2020pre} $I(\omega)\sim \omega^5$  for $N = 32768$ that is much larger than $N_c$ in very poorly annealed IPL model glasses quenched from a melt. This attracts us to test directly whether the $I(\omega)\sim \omega^5$ scaling  could still work at much lower frequencies in this $N = 32768$ system.
Following nearly the same simulation details as used in Ref.~\cite{Lerner2020pre}, we generated very poorly annealed IPL model glasses with $N=32768$, and  the corresponding $I(\omega)$ has been shown in Fig.~\ref{fig1}(c). Unexpectedly, for $N=32768 > N_c$, one could still observe an obvious  deviation of the low-frequency data from the $I(\omega)\sim \omega^5$  scaling.
However,  the possibility that the low-frequency deviation  is due to finite-size effects still can not be  ruled out. If there is a critical system size above which our observed low-frequency deviation disappears, we would anticipate from our results that the critical size should be much larger than the size determined from  the core size
 of soft quasi-localized modes~\cite{Lerner2020pre}.
 However, given the large ensemble sizes needed in simulations to probe very low frequencies, it could be very difficult to determine how large the largest simulated system  should be when checking finite-size effects.  Hence, we expect future theoretical work could shed some light on this issue.


\begin{figure}
\centering
\includegraphics[width=0.475\textwidth]{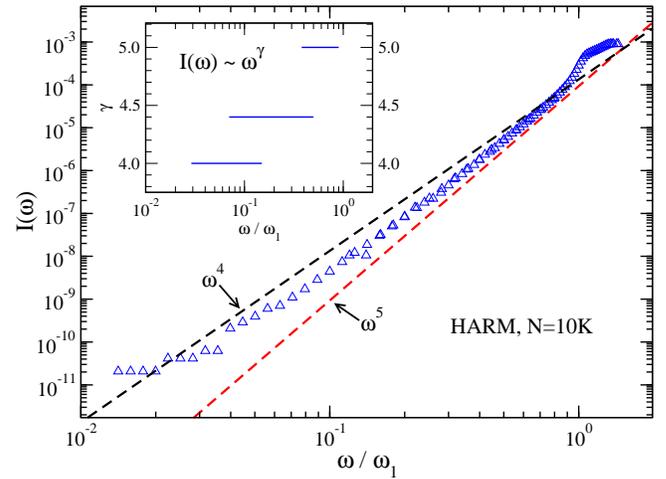}
\caption{\label{fig4} Frequency $\omega/\omega_1$ dependence of the cumulative density of states  $I(\omega)$ in 3D HARM  model glasses with $N=10K$. The first sound mode frequency $\omega_1\approx 0.10$.
The red and black dashed lines represent power laws of $\omega^5$ and $\omega^4$, respectively. It's obvious that the $\omega^5$ scaling breaks down at very low frequencies. (Inset) The scaling exponent $\gamma$ versus the estimated corresponding frequency regime where the plot of $I(\omega)/\omega^{\gamma}$ exhibits  a  plateau.
}
\end{figure}

In addition, we  also studied $I(\omega)$ in a small  HARM model system with $N=512$ besides  the one with $N=10K$ as shown in Fig.~\ref{fig1}(a), and the comparison of $I(\omega)$ in the two systems  is shown in Fig.~\ref{fig3}. Within an overlapping  frequency region marked with a horizontal line with arrows in both ends in Fig.~\ref{fig3}(a), one can observe there is  an  upturn to the lower frequency end in the $I(\omega)/\omega^5$ plot for $N=512$, suggesting $\gamma<5$ in $I(\omega)\sim \omega^\gamma$; however, in the same frequency range, $\gamma=5$ in the $N=10K$ system since there is a reasonable plateau in the reduced plot. As shown in Fig.~\ref{fig1}(a) and again here, the $I(\omega)\sim \omega^5$ scaling breaks down at lower frequencies in the $N=10K$ system, which motivates us to see whether the $I(\omega)$ data in  $N=512$ and $N=10K$ may share the same scaling law of  $\omega$ in some frequency range. Interestingly, we could identify  a frequency region where the scaling of  $I(\omega)\sim \omega^\gamma$ with $\gamma=4.4$ can work reasonably in the two systems, see Fig.~\ref{fig3}(b).  Therefore, we conclude  whether $\gamma$ in $I(\omega)\sim \omega^\gamma$ in a small system is the same as  in  a large one  seems to depend on the frequency range examined.


\begin{figure*}
\centering
\includegraphics[width=0.8\textwidth]{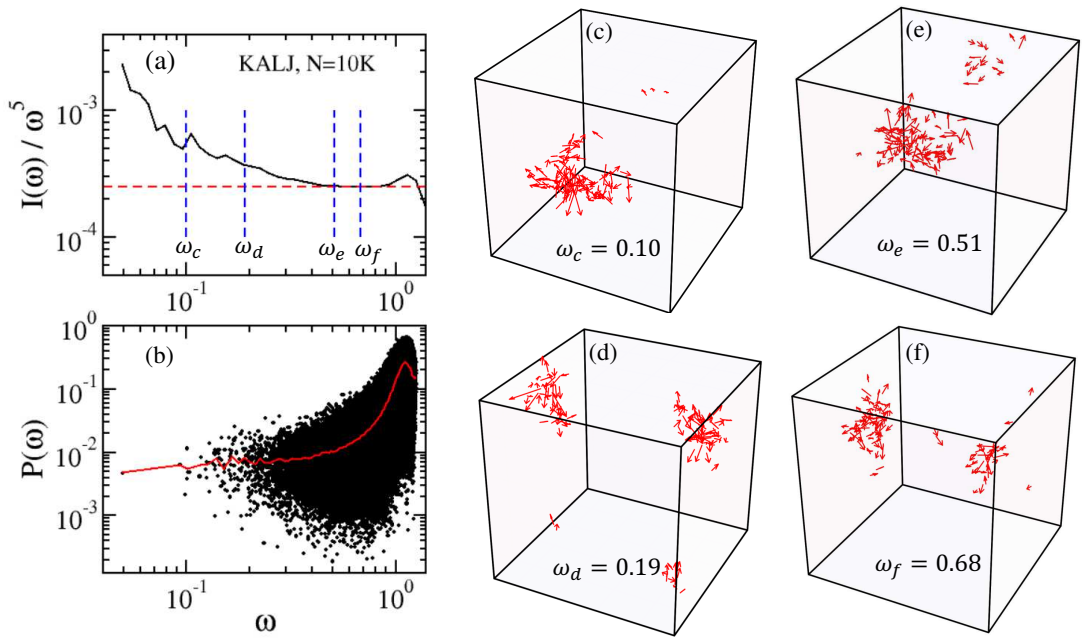}
\caption{\label{fig5} (a) The reduced cumulative density of states  $I(\omega)/\omega^5$ in very poorly annealed 3D KALJ  model glasses with $N=10K$. The vertical dashed lines indicate the frequencies of the modes in from panel (c) to panel (f), and the horizontal dashed line is a guide to eyes.
 (b) Scatter plot of participation ratio $P(\omega)$ versus $\omega$. Each point corresponds to $P(\omega)$ in one mode with $\omega$. The red line represents the average participation ratio in different frequency bins for the same data employed to make the scatter plot of $P(\omega)$. For reference, the ensemble size $N_{En}=99.2K$ in (a), and $N_{En}=33.4K$  in (b).
(c)-(f): 3D visualization  of low-frequency eigenvectors. The polarization vectors of particles are denoted by arrows, and only those polarization vectors  whose lengths are among the  1\% largest  in each configuration  are shown.
The four modes  come from the lowest-frequency modes of four different configurations.
}
\end{figure*}

 Though we do not seek in this work to claim what the value of the scaling exponent $\gamma$ in $I(\omega)\sim \omega^\gamma$ is on approaching zero frequency, we tried to make a tentative estimate of $\gamma$ in the low-frequency regimes available to us. Figure~\ref{fig4} shows the not-reduced cumulative density of states  in the HARM model glasses with $N=10K$.  We indicate two power law lines of $\omega^5$ and $\omega^4$ as a guide to eyes.  It's visible the $\omega^5$ scaling could not work   even in the  $I(\omega)$ versus $\omega$ plot in the lowest-frequency regime, while  the $\omega^4$ scaling seems to work in this regime. To be more quantitative, we show in the inset three values of  $\gamma$, i.e., $\gamma=5$, $\gamma=4.4$, and $\gamma=4$,  versus the  associated frequency regime where the reduced plot, $I(\omega)/\omega^{\gamma}$, exhibits a  plateau. One could observe $\gamma=5$ is valid in an intermediate-frequency regime, whereas the scaling with $\gamma=4$ is a good  description to our lowest-frequency data.

 We go a little further to characterize the modes' eigenvectors  to see whether there is a visible difference between the characteristics of modes in the frequency regime where $I(\omega)\sim \omega^{5}$ works, denoted in the following  as the $\omega^5$ regime, and modes in the frequency regime where $I(\omega)\sim \omega^{5}$ does not work, denoted as the $\omega^5$ breakdown regime. Figure~\ref{fig5}(a) shows $I(\omega)/ \omega^{5}$ against $\omega$ and  serves mainly here as a guide. We show the 3D visualization of four eigenvectors in Figs.~\ref{fig5}(c), (d), (e) and (f). For visualization purpose, we only show  particles' polarization vectors whose lengths are no smaller than the largest 1\%.
 The frequencies of the four modes visualized are marked with vertical dashed lines in Fig.~\ref{fig5}(a). Specifically, the two modes in Figs.~\ref{fig5}(c) and (d) are in the $\omega^5$ breakdown regime, whereas  the two modes in Figs.~\ref{fig5}(e) and (f) are in  the $\omega^5$  regime. We could observe the four modes all feature disordered cores decorated by  particles with different vibrational displacements, which is the typical characteristic of quasi-localized modes~\cite{lerner_prl2016,Shimada2018pre}. However, we could not distinguish the four  modes purely from the spatial characterization performed here.

 Hence, we made a more quantitative characterization of low-frequency modes by calculating the participation ratio $P(\omega)$ which measures the degree of the mode localization.  A more localized mode usually has a smaller value of  $P(\omega)$. $P(\omega)=1/N$ if there is only  one particle participating in the  vibration of one mode. However, the visible fluctuation in the scatter-plot of $P(\omega)$ in Fig.~\ref{fig5}(b) makes it hard to determine  how the mode localization evolves with decreasing frequencies. Moreover, we could observe  participation ratios for modes at similar frequencies may  differ markedly.
 We notice it was argued in Ref.~\cite{Richard2020prl} that  the average participation ratio $P_{\rm Ave}(\omega)$ is nearly independent of frequencies in the  $\omega^5$ regime, which stimulates us to check how $P_{\rm Ave}(\omega)$ evolves with $\omega$  in the $\omega^5$ breakdown regime. We calculated $P_{\rm Ave}(\omega)$, see the red line in Fig.~\ref{fig5}(b).   Interestingly,  it seems that $P_{\rm Ave}(\omega)$  exhibits   mild dependence on $\omega$ as well even in the $\omega^5$ breakdown regime. Therefore, our  results in Fig.~\ref{fig5} suggest that modes in  the $\omega^5$ regime and modes in the $\omega^5$ breakdown regime could have similar structures and localizations.  However, we do not exclude that a more sophisticated characterization of  these modes may reach a difference.


\section{CONCLUSION and Discussion}

In summary, we examined the density of excess modes at ultra-low frequencies far below the first sound mode frequency $\omega_1$ in 3D glasses. We find the  previously reported $D(\omega)\sim \omega^{4}$ scaling below $\omega_1$  could not work all the way to very low frequencies, though it may work  in an intermediate-frequency regime in some glasses.
Moreover, the breakdown of the $\omega^{4}$ law could be observed in  model glasses with different potentials as well as in glasses with small difference in the glass stability.
 We notice that the  $D(\omega)\sim \omega^{4}$ law which is assumed to extend to zero frequency below $\omega_1$, is dominant  in current simulation studies of 3D small glasses.
 Our work  thus makes a  numerical attack on the validity of the quartic law at very low frequencies below $\omega_1$ in 3D glasses.

Our results  imply  that there may be still a long way to go to have a comprehensive knowledge of the  low-frequency density of  excess modes below $\omega_1$ in 3D glasses.
We notice that one recent study~\cite{Stanifer2018pre} using a simple random matrix model  found a $D(\omega)\sim \omega^{3}$ scaling at very low frequencies, and a $D(\omega)\sim \omega^{4}$ scaling in an intermediate-frequency regime, which is, to some extent, analogous to our numerical observation. We also notice that Krishnan, Ramola,  and Karmakar~\cite{Krishnan2D}, found  low-frequency modes below $\omega_1$ exhibit a density of states $D(\omega)\sim \omega^{5}$ in shear-stabilized  structural glasses, which  suggests the breakdown of the $\omega^{4}$ law though in a different context. More work is needed to reconcile different results regarding the breakdown of $D(\omega)\sim \omega^{4}$  below $\omega_1$.
  Due to  the need of a very large amount of computation,  it is now impossible for us to determine precisely how  $D(\omega)$ scales with $\omega$ at very low frequencies below $\omega_1$, but future work may be able to resolve this.

   We propose several factors which could be taken into account in future studies.  First,
   it would be  necessary to check whether increasing the glass stability spectacularly  could change our major conclusion in this work,  because recent  studies~\cite{Wang2019NC,Wang2019SMattenuation,Flenner_SM2020,Lerner2020pnas,lerner2019JCP,Miyazakistability,Ozawa2018PNAS,hellman,specific_heat_pnas,Parisi2019prl,Alisoftmatter} suggest the glass stability  matters a lot in modulating various glassy properties.
   Second, it would be interesting to investigate how $D(\omega)$ behaves at frequencies far below $\omega_1$ in large spatial dimensions, e.g., in 4D. A compilation of the results in 3D glasses reported in this work and the results in 2D glasses reported in Ref.~\cite{Wang2021prl}  suggests the quartic frequency dependence of $D(\omega)$ below $\omega_1$ breaks down at very low frequencies in both 2D and 3D glasses. In 4D glasses, it was argued~\cite{Lerner2Dprl} that   $D(\omega) \sim \omega^4$ below $\omega_1$, but it's unknown whether the quartic scaling could persist at much lower frequencies. For glasses in very large spatial dimensions, we  notice that it has been  demonstrated~\cite{Shimada2020pre} numerically the quadratic scaling, $D(\omega) \sim \omega^2$, predicted by mean-field theories~\cite{meanfield1,meanfield2} could work well at very low frequencies. Finally, it has been shown properties of vibrational modes in   glasses near the jamming transition~\cite{OHern2003} are fascinating~\cite{Shimada2018pre}, and it may be  intriguing to extend the numerical investigation of vibrational modes in nearly jammed glasses to much lower frequencies below $\omega_1$. We note all our glasses examined in this work are  far away from the jamming transition.

In the end,  we reiterate that our main aim in  this work is to show our numerical observation of the breakdown of the  quartic scaling of the density of low-frequency  excess modes below the first sound mode.  We  also reiterate that we  followed the commonly used procedure to generate glasses and then studied low-frequency modes below the first sound mode, and our major difference than previous studies is that we probed lower frequencies by using  extremely large ensemble sizes. We note that it would be very important to check whether our new results at very low frequencies are due to some hidden factors; these factors may come into play or even dominate at very low frequencies, but  are not taken into account when doing computer simulations. We believe this line of research may attract some interesting studies.

\section*{ACKNOWLEDGMENTS}

We thank Elijah Flenner, Grzegorz Szamel, and Ding Xu for suggestions/discussions, and Edan Lerner for comments on our major results.
L. Wang and L. Fu acknowledge the support from   National Natural Science Foundation of China (Grant No. 12004001), Anhui Projects (Grant No. S020218016 and Grant No. Z010118169), Hefei City (Grant No. Z020132009), and  Anhui University (start-up fund). Y. Nie acknowledges the support from China Postdoctoral Science Foundation (Grant No. 2021M690328). We  also acknowledge Hefei Advanced Computing Center, Beijing Super Cloud Computing Center, and the High-Performance Computing Platform of Anhui University for providing computing resources.

\end{document}